# Some features of anomalous conductivity of vinyl and vinyl chloride copolymer films

V.I.Kryshtob, V.F.Mironov, L.A.Apresyan, D.V.Vlasov, S.I. Rasmagin, T.V. Vlasova

Prokhorov General Physics Institute, Russian Academy of Science
ul.Vavilova 38, Moscow, 119991 Russia

ABSTRACT

Manifestations of anomalous conductivity in polar dielectric films, is demonstrated for the first time on samples of copolymer of vinylene and vinyl chloride, as was previously observed on modified PVC samples. On the base of experimental results important new insights into the physical sense of traditionally used in these conditions specific resistivity indicators both volume and surface should be replaced by transverse and longitudinal resistance (according to polymer film surface) respectively, because specific resistivity of polymer composite does not permits to calculate as usual resistance of sample of arbitrary form.

ВВЕДЕНИЕ

Изменение электропроводящих свойств различных полимеров способами их физико-химической модификации введением в их объем, например, различных ингредиентов (наполнителей, пластификаторов, стабилизаторов и т.д.) известно давно [1,2]. С другой стороны, изменять в достаточно широких пределах электропроводящие свойства полимерного материала можно с помощью выбора и использования того или иного вида самого полимера. При этом наиболее высокими электропроводящими свойствами обладают полимеры с системой полиеновых сопряженных связей (ПСС) в цепи самой макромолекулы (это наиболее известные сегодня полипиррол, полипарафенилен, политиофен, полиацетилен и т.д. [1,3]). Очень часто для усиления электропроводящих свойств полимеров, содержащих в цепи ПСС, дополнительно используют способы их химической модификации (допирование) за счет реакции полимера с донорами или акцепторами электронов, приводящих зачастую к чрезвычайно масштабным изменения электропроводности полимера, вплоть до перехода их в металлической состояние [1]. Однако, во всех вышеперечисленных случаях после изменения электропроводности полимерного материала методами физико-химической или химической модификации характер электропроводности, как правило, носит устойчивый, подчиняющийся закону Ома характер.

При этом вполне понятно, что в силу имеющихся больших возможностей масштабного изменения электропроводности в полимерах, содержащих в своем составе ПСС, основной интерес исследователей обращен именно к ним. Однако основным тормозом при их практическом использовании является практически их полная технологическая непригодность [1-3].

С этой точки зрения наиболее перспективным и логичным являлось бы использование одного из самых технологичных и массово используемых полимеров в мире, таких например, как ПВХ.

Однако последний, обладая чрезвычайно низкой электропроводностью, и не содержащий при этом в составе своих макроцепей ПСС, является типичным представителем класса диэлектриков (удельное объемное сопротивление $10^{13}$- $10^{16}$Ом.см)[2]. Но, с другой стороны, хорошо известно, что полиацетилен (ПАц), содержащий в составе своих макромолекул исключительно ПСС и (условно) представляющий собой ПВХ, подвергнутый операции 100%-дегидрохлорирования, обладает значительно большей электропроводностью (удельное объемное сопротивление $10^8$-$10^{10}$ Ом.см) хотя абсолютно непригоден при этом с технологической точки зрения [1].

Учитывая эти обстоятельства, было решено путем частичного термического дегидрохлорирования исходного ПВХ попытаться отыскать оптимум электропроводящих и технологических свойств сополимеров винилхлорида и винилена.

С этой целью в данной работе исходные образцы ПВХ подвергались химической модификации методом термолиза последнего (без использования при этом каких-либо пластификаторов, специальных добавок, ионизирующего излучения и т.д.)[4].

## ЭКСПЕРИМЕНТАЛЬНАЯ ЧАСТЬ

Опытные образцы частично дегидрохлорированного ПВХ методом термолиза в растворе получали следующим образом. Вначале получали 4% раствор ПВХ (марки С-70) в растворителе (ацетофеноне). Растворение ПВХ осуществляли при перемешивании при комнатной температуре в течение 12 часов до получения гомогенного прозрачного раствора. В дальнейшем раствор помещался в пробирку и подвергался термолизу при $T=190^0C$ в течение 20-480минут.

Порция , необходимая для получения образца пленки заливалась на стеклянную подложку и подвергалась сушке при $T=95^0C$ в термошкафу в течение 48 часов. Далее полученные образцы подвергались визуальному, технологическому и органолептическому контролю (прозрачность, цвет, легкость снятия с подложки, прочность, изгибостойкость, залипаемость и т.д.).

Измерения образцов ПВХ-пленок по показаниям электропроводности осуществляли на приборе Е6-3, описанном в [5-11] с ГОСТированной кольцевой измерительной ячейкой, а также с использованием микронаноамперметра Ф136.

Важно отметить, что при использовании конфигурации определения значений продольного (поверхностного) сопротивления образцов сополимерных пленок ($R_s$) [4], нижняя (не соприкасающаяся с измерительными электродами) поверхность пленки образца была изолирована пленкой из фторопласта.

## ОБСУЖДЕНИЕ РЕЗУЛЬТАТОВ

Данные по основным технологическим и электрофизическим свойствам полученных в лабораторных условиях образцов исходного ПВХ и частично дегидрохлорированых его образцов (сополимеров винилхлорида и винилена) представлены в Таблице 1.

| № образца | Время дегидро-хлорирования, мин | Толщина пленок, мкм | Внешний вид и органолептические свойства | Поперечное (объемное) сопротивление, $R_V$, Ом | | Продольное (поверхностное) сопротивление, $R_S$, Ом | |
|---|---|---|---|---|---|---|---|
| | | | | СНП | СВП | СНП | СВП |
| (1) | (2) | (3) | (4) | (5) | (6) | (7) | (8) |
| №1 (исх. ПВХ) | 0 | 14 | Бесцветная, прозрачная, хрупкая, ломающаяся пленка | $>10^{12}$ | Переход в СВП не набл. | $>10^{12}$ | Переход не наблюдается |
| №2 | 20 | 12 | Прозр., менее хрупкая пленка слабо-желтого цвета | $>10^{12}$ | Переход в СВП не набл | $>10^{12}$ | Переход не наблюдается |
| №3 | 240 | 10 | Прозр., желтого цвета, прочная, незалипающая, гнущаяся, легко снимающаяся со стекл.подл. пленка | $8,6 \cdot 10^{11}$ | 1,3 | $>10^{12}$ | Переход не наблюдается $2,2 \cdot 10^4$ |
| №4 | 320 | 10 | Прозр., желтого цвета, прочная, незалипающая, гнущаяся, легко снимающаяся со стекл.подл. пленка | $7,1 \cdot 10^{11}$ | 0,6 | $>10^{12}$ | Переход не наблюдается $1,8 \cdot 10^{11}$ |
| №5 | 480 | 11 | Прозр., более насыщенного желтого цвета, незалипающая, легко снимающаяся со стекл.подл., гнущаяся пленка | $4,3 \cdot 10^{11}$ | 0,5 | $>10^{12}$ | Переход не наблюдается |

Органолептический и визуальный анализ полученных лабораторных образцов на базе ПВХ однозначно свидетельствует, что появление в ПВХ в процессе его термолиза ПСС не только не ухудшает его технологических свойств (по сравнению с исходным ПВХ и тем более с полиацетиленом) но способствует даже их некоторому улучшению.

Но наиболее впечатляющим оказалось то обстоятельство, что для части образцов сополимеров винилхлорида и винилена (обр. № 3-5) переход из СНП в проводимости СВП носит столь масштабный характер (более 12 порядков)., что образец из состояния типичного диэлектрика может переходить в разряд проводников.

Отметим также спонтанный характер этих переходов. Причем переход в СВП частично дегидрохлорированных образцов ПВХ осуществлялся гораздо легче и быстрее в случае большего времени термолиза (т.е. в условиях предположительно большей степени их дегидрохлорирования).

Следует особо отметить, что переходы из СНП в СВП наблюдались лишь в случаях измерения поперечного сопротивления ($R_v$) и оставались практически без изменения при измерениях продольного сопротивления образцов ($R_s$), что лишний раз может свидетельствовать об аномальных (не подчиняющихся закону Ома) изменениях электропроводности частично дегидрохлорированных образцов ПВХ. Кроме того, экспериментально установлено, что для образцов сополимера изменения давления, температуры и напряжения. Сопровождались достаточно заметными изменениями сопротивления, что предполагает в дальнейшем их более детальные исследования этих воздействий.

Таким образом, с учетом вышеизложенного можно сделать следующие основные выводы:

1) Впервые в образцах сополимера винилена и винилхлорида, удалось обнаружить аномальный характер проводимости и переходы из СНП в СВП, носящие обратимый, спонтанный характер.

2) Масштаб изменений образцов в СНП иСВП носил при этом гигантский характер (более 12 порядков), т.е. позволял образцам переходить из состояния диэлектрика в состояние проводника,

3) Абсолютное значение максимального сопротивления для сополимера винилхлорида-винилена в СВП (образец N5) Значительно (на несколько порядков) превосходило значение проводимости самого (100%) ПАц, а по комплексу физико-механических и технологических свойств - исходного ПВХ.

4) Близкие к металлическим, измеренные значения сопротивления $R_v$ при использовании стандартной формулы $R=\rho L/S$, дают в пересчете на геометрию $R_s$, значение сопротивления порядка кОм. Однако, в эксперименте это сопротивление превышает $10^{12}$ Ом, что однозначно свидетельствует о том, что расчет удельных сопротивлений и соответствующей формулы пересчета ($R=\rho L/S$) для полимерных нанокомпозитов не имеет физического смысла даже для оценок по порядку величины.

.